\documentclass[reprint,aps,amsmath, amssymb]{article}
\usepackage{geometry}                
\usepackage[english,russian]{babel}
\usepackage{graphicx}
\usepackage{amssymb}
\usepackage{amsmath}

\renewcommand{\vec}[1]{{\mbox{\bf{#1}}}}

\newcommand{\PDEV}[2]{{\frac{\partial{#1}}{\partial{#2}}}}

\newcommand{\PD}[2]{{\frac{\partial^2{#1}}{\partial{#2}^2}}}
\newcommand{\SCAL}[2]{(\vec{#1}\!\cdot\!\vec{#2})}
\newcommand{\rme}{\mathrm{e}}
\newcommand{\rmi}{\mathrm{i}}

\newcommand{\rmz}{\mathrm{z}}

\begin{document}
\title{Advances in the studies of anomalous diffusion in velocity space}
\author{A.A. Dubinova, S.A. Trigger}
\date{Eindhoven University of Technology, Eindhoven, The Netherlands;\\
Joint Institute for High Temperatures RAS, Moscow, Russia}                                           

\maketitle

\section*{Abstract}
	A generalized Fokker-Planck equation is derived to describe particle kinetics
	in specific situations	when the probability transition function (PTF) has a long
	tail in momentum space.
	The equation is valid for an arbitrary value of the transferred in a collision act
	momentum and for the arbitrary mass ratio of the interacting particles.
	On the basis of the generalized Fokker-Planck equation anomalous diffusion
	in velocity space is considered
	for hard sphere model of particle interactions, Coulomb collisions and interactions
	typical for dusty plasmas.
	The example of dusty plasma interaction is peculiar in way that it leads to a new
	term in the obtained Fokker-Planck-iike equation due to the dependence of the
	differential cross-section on the relative velocity.
	The theory is also applied to diffusion of heavy particles in the ambience of light particles
	with a prescribed power-type velocity distribution function. In general, the theory is applicable to consideration of anomalous diffusion in velocity space if the typical velocity of one sort of particles undergoing
	diffusion is small compared to the typical velocity of the background particles.\\

PACS number(s): 52.27.Lw, 52.20.Hv, 05.40.-a, 05.40.Fb

\section*{Introduction}

	Interest to anomalous diffusion in coordinate space is explained by its great variety
	of applications, i.e., semiconductors, polymers, some granular systems,
	plasmas under specific conditions, various objects in biological
	systems, physical-chemical systems, and others.

	Non-linear time dependence $<r^2(t)>$ of the mean-squared
	 displacement has been experimentally observed,
	in particular, under essentially non-equilibrium conditions and
	in some disordered systems \cite {2,3}.
	As is well known there are two types of anomalous diffusion in coordinate space
	which are referred to as superdiffusion
	$<r^2(t)>\sim t^\alpha$ $(\alpha>1)$ and subdiffusion $(\alpha<1)$ \cite {4}.
	For description of these two anomalous diffusion regimes a number of
	efficient models have been proposed.
	The continuous time random walk (CTRW) model by Scher and Montroll \cite {5}
	for the process of subdiffusion is a basis for understanding photoconductivity
	in strongly disordered and glassy semiconductors.
	The Levy-flight superdiffusion model \cite {6} describes such phenomena as
	self-diffusion in micelle systems \cite {7} as well as reaction and transport in
	polymer systems [8].  Another application of the model is studying stochastic
	behavior of financial market indices \cite {9}.
	The method of so-called fractional differential equations in coordinate and time
	spaces has been successfully developed to treat both cases of anomalous diffusion
	 \cite {10}.

	Recently a more general approach has been proposed in \cite {11} which not only
	reproduces the results of the standard fractional differentiation method (whenever
	it is applicable) but also allows to tackle more complicated cases of anomalous
	diffusion. In \cite {12}, for example, this approach is used for studying diffusion
	in a time-dependent external field.

	Some aspects of anomalous diffusion in velocity space have
	been considered in papers \cite {13}-\cite{ 18}. However, compared to anomalous
	diffusion in coordinate space anomalous diffusion in velocity space is still poorly
	studied but it is actively attracting more  and more attention
	 \cite {Trigger 2009}-\cite {TrEbHeScSok 2010}.

	In this paper the problem of anomalous diffusion in momentum
	(velocity) space is considered in detail in the spirit of the approach
	proposed in \cite{11} for the diffusion in coordinate space.
	Extra terms (cp. \cite {Trigger 2009}-\cite {TrEbHeScSok 2010}) of the generalized 				Fokker-Planck equation are found. The results are applied to hard sphere model of collision,
	Coulomb systems and dusty plasma. In dusty plasma the additional terms are numerically
	small but not negligible due to
	to velocity dependence of the cross-section.

	The paper is organized as follows. Diffusion in velocity space
	for the cases of normal and anomalous behavior of the PTF
	is presented in Section 1. Starting from the Boltzmann-type PTF
	we arrive at explicit expansion PTF into a series which is applicable not only to the
	Boltzmann-type processes but also to a wide class of processes
	with other types of PT-functions. The generalized Fokker-Planck equation is
	presented in Section 2.
	Special cases of anomalous diffusion with the specific power-type
	distribution function of ambient light particles are analysed in Section 3.
	In this section the cases of hard sphere and Coulomb models of diffusion
	are considered.
	Subsection 3.3 is devoted to the example of anomalous diffusion in dusty plasma
	where the differential cross-section depends on velocity of the colliding particles.
	In this case the additional term in the generalized kinetic equation is not negligible.

\section{Diffusion in velocity space on the basis of the master-type equation}

	Let us consider diffusion in velocity space using an approach
	based on the corresponding master equation for the distribution function $f({\bf p},t)$
	which describes the balance of particles coming at and from the point $\vec{p}$ at the instant $t$.
	The master equation written in velocity space reads (see, e.g., \cite{Lifshitz1979})

\begin{eqnarray}
	&\frac{df({\bf p},t)}{dt} = \int d{\bf q} \left\{W ({\bf q, p+q})
	f({\bf p+q}, t)\right.\nonumber \\
	&\left.- W ({\bf q, p}) f({\bf p},t) \right\}.
				\label{DC2b}
\end{eqnarray}
	The probability transition function $W({\bf q, p'})$ describes the probability that a particle with
	momentum ${\bf p'}$ passes from the point
	${\bf p'}$ in the velocity phase space to the point ${\bf p}$ per unit time by transferring
	momentum ${\bf q=p'- p}$ to the surrounding medium. Under the assumption that
	the characteristic transfered momentum ${\bf q}$ is much smaller than
	${\bf p}$ equation (\ref{DC2b}) can be expanded into a series with respect
	to ${\bf q}$ to the second order.
	The expansion leads to the usual form of the Fokker-Planck
	equation for the density distribution function $f({\bf p},t)$

\begin{eqnarray}
	&\frac{df({\bf p},t)}{dt} = \frac {\partial}{\partial p_\alpha}
	\left[ A_\alpha ({\bf p}) f({\bf p},t) \right.\nonumber\\
	&\left.+ \frac{\partial}{\partial p_\beta} \left(B_{\alpha\beta}({\bf p}) f({\bf p},t)\right)\right],
				\label{DC3b}
\end{eqnarray}
	where
\begin{eqnarray}
	&A_\alpha({\bf p}) = \int  q_\alpha W({\bf q, p})d\vec{q},\\
	&B_{\alpha\beta}({\bf p})= \frac{1}{2}\int q_\alpha q_\beta W({\bf q, p})d\vec{q}.
			 	\label{DC4b}
\end{eqnarray}
	The coefficients $A_\alpha$ and $B_{\alpha \beta}$ are responsible for the processes of
	friction and diffusion, respectively. The subscripts $\alpha$ and
	$\beta$ correspond to the coordinate axes.

	For the Boltzmann elastic collisions
	the PTF-function $W({\bf q, p)}$ has been found in \cite{Hoare1971,Nielsen1964} and can be represented in the form \cite{11}
\begin{eqnarray}
	W({\bf p, q})= \frac{1}{\mu^2} \int d{\bf u}\, \delta \left(
	{\bf u\!\cdot\! q}+ \frac{q^2}{2\mu} \right)\nonumber\\
	\times \frac{d \sigma}{do} \left
	[\arccos\, (1-\frac{q^2}{2\mu^2 u^2}), u \right] f_b (\bf u+ v).
				\label{w_b}
\end{eqnarray}
	Here ${\bf p}=M{\bf v}$ is the momentum of a scattered particle,
	$\mu=Mm/(m+M)$ is reduced mass
	and $f_b$ is the distribution function
	for the background scattering particles.
	The relative collision velocity is denoted with $\vec{u}$ and
	$d \sigma (\chi,u)/ do$ is the scattering differential cross-section
	calculated in the coordinate system in which a scattering particles with mass $m$ is fixed
	throughout the whole act of collision \cite{Chapmen1999, Liboff1969}.
	The scattering angle $\chi = \arccos[1-q^2/(2\mu^2u^2)]$ stands for the rotation of
	the relative velocity $\vec{u}$.
	Assuming that the vector $\vec{q}$ is directed along the axis $\vec{z}$ and performing
	integration with respect to the projection of $u$ on the axis $\vec{z}$
	one arrives at the more explicit form of the equation for the PTF-function
\begin{equation}
	\begin{split}
	&W({\bf q, p}) = \frac{1}{\mu^2 q} \int d\vec{u}_{\perp}\frac{d \sigma}{do}
	\left[\chi, u \right]_{ u^2= u_\rmz^2+u_{\perp}^2}\\
	&\times f_b \left[(\vec{u}_{\perp}+\vec{v}_{\perp})^2+u_\rmz^2-\frac{\bf q \!\cdot \!v}{\mu}
	+\frac{(\vec{q}\!\cdot\!\vec{v})^2}{q^2}\right]\!,
	\end{split}
				\label{DC13z}
\end{equation}
	where $u_\rmz=-q/2\mu\,$, $\vec{u}_{\perp}$ and $\vec{v}_{\perp}$ mean projections of the vectors
	$\vec{u}$ and {\vec{v}}, respectively, on the plane $\vec{xy}$
	perpendicular to the axis $\vec{z}$.

	For the scattering cross-sections independent on $u$ (e.g., for hard sphere
	or Coulomb interactions)
	equation (\ref{DC13z}) is reduced to the following equation by changing the variable of integration
	$\tilde{\vec{u}}=\vec{u}_{\perp}+\vec{v}_{\perp}$
\begin{equation}\begin{split}
	&W({\bf q, p}) =
	\frac{\pi}{\mu^2 q}\frac{d\sigma}{do}\\
	&\times\int f_b \left[\tilde{\vec{u}}^2+\frac{q^2}{4\mu^2}-\frac{\bf q\! \cdot \!v}{\mu}\right.
	\left.+\frac{(\vec{q}\!\cdot\!\vec{v})^2}{q^2}\right]d\tilde{\vec{u}}^2\!,
			\end{split}
				\label{DC14z}
\end{equation}

	Now let us consider the case when the characteristic velocity of particles diffusing in
	velocity space is small compared to the characteristic velocities in the
	surrounding medium (e.g., \cite{Lifshitz1979}). For the Boltzmann-type
	collisions it means that there are two types of particles -- heavy
	particles undergoing diffusion with the characteristic velocity $ v\sim p/ M$
	and light particles characterized by velocity $u$ and $u \gg v$.
	In the zero order approximation with respect to $\vec{p}$
	when we omit any dependence of the PTF on $\vec{p}$ (i.e., $W(\vec{q},\vec{p})=W(\vec{q},\vec{p}=0) = W(q)$)
	the coefficients $A_{\alpha}$ and $B_{\alpha\beta}$ yield the equations
\begin{equation}
	\begin{split}
	&A_{\alpha}=0,\\
	&B_{\alpha\beta} \equiv \delta_{\alpha\beta}B
	= \frac{\delta_{\alpha\beta}}{6}\int q^2 W(q)d\vec{q},
	\end{split}
				\label{DC6b}
\end{equation}
	where $\delta_{\alpha\beta}$ is the Kronecker delta.
	The neglect of $\vec{p}$-dependence is not correct. Usually the coefficient $A_\alpha$
	for the Fokker-Planck equation is found assuming that
	the stationary distribution function is Maxwellian.
	This assumption leads to the standard relation for the coefficients $MT
	A_\alpha(p)=p_\alpha B$ where $M$ is mass of the diffusing particle
	and $T$ is temperature of the particles in equilibrium.
	The relation is analogous to the Einstein
	relation in coordinate space.

	For the systems far from equilibrium, for example, for slowly
	decreasing PTF, the approach above is not
	applicable. In this case rigorous calculation of the coefficients $A_{\alpha}$
	and $B_{\alpha\beta}$ requires the next order
	of approximation with respect to the smallness of $\vec{p}$ \cite{Trigger 2009}.
	As the function $W(\bf {q,p})$ is a scalar it can be regarded as
	a function depending only on variables $q,\: {\bf q\! \cdot\! p},\: p^2$.
	We imply that the PTF is such a function of $\vec{q}\cdot\vec{p}$ and $p^2$ that it
	can be analytically expanded into a series in the vicinity of the point
	$\vec{q}\cdot \vec{p}=0$  $p^2=0$. Note that in the expansion of the PTF we leave only
	those terms which have $p$ to the power no higher than 2.
	Expansion of $W(\bf {q,p})$ into a series to the second order is given by
\begin{equation}
	\begin{split}
	&W({\bf q,p)}\simeq W(q)+\tilde W'(q)({\bf q\! \cdot\! p})\\
	 &+\frac{1}{2}\tilde W''(q) ({\bf q\! \cdot\! p})^2+\hat{W}'(q)p^2\!\!,
	 \end{split}
				\label{DC7b}
\end{equation}
where
\begin{equation}
	\begin{split}
&	\tilde W'(q)= \left.\PDEV{W (q, {\bf q \!\cdot \!p}, p^2)}{(\bf q\!\cdot\! p)}
	\right|_{{\bf q \cdot p}=0, \,p^2=0}\!\!\!\!\!\!\!\!\!\!\!\!\!\!\!\!\!\!\!\!\!\!\!\!\!\!\!\!,\\
&	\tilde W''(q) =\left.\PD{W (q, {\bf q\! \cdot\! p},p^2)}{(\bf q\!\cdot \!p)}
	\right|_{{\bf q \cdot p}=0,\, p^2=0}\!\!\!\!\!\!\!\!\!\!\!\!\!\!\!\!\!\!\!\!\!\!\!\!\!\!\!\!,\\
&	\hat W'(q)= \left.\PDEV{W (q, {\bf q\! \cdot\! p}, p^2)}{p^2}\right|_{{\bf q \cdot p}=0, \,p^2=0}
	\!\!\!\!\!\!\!\!\!\!\!\!\!\!\!\!\!\!\!\!\!\!\!\!\!\!\!\!.
	\end{split}
\end{equation}
	 Basically, the developed approximation is true if the typical velocity of one sort of particles undergoing
	diffusion is small compared to the typical velocity of the background particles.
	Let us note that for the differential cross-sections independent on the
	relative velocity $u$ the coefficient $\hat W'(q)\equiv0$ as it can be seen
	 from equation~(\ref{DC14z}).
    	In general, this term is not negligible and it should be taken into account (it was omitted in
    \cite{Trigger 2009, TrEbHeScSok 2010}).

	Then with the necessary accuracy we find the coefficient $A_\alpha$
\begin{equation}
	\begin{split}
	&A_\alpha({\bf p}) = \int   q_\alpha q_\beta p_\beta \tilde W'(q)d\vec{q}\\
	&	= p_\alpha \int q_\alpha q_\alpha  \tilde W'(q) d\vec{q}\\
	&	=\frac{p_\alpha}{3} \int  q^2  \tilde W'(q)d\vec{q}.
		\end{split}
			\label{DC10b}
\end{equation}
	Remarkably, if the equality $\tilde W'(q)= mW(q)/ (2M\mu T)$
	is fulfilled for the function $W({\bf q,p)}$
	the following relation for the coefficients $A_\alpha$ and $B$
	takes place
\begin{equation}
	\frac{\mu}{m} MT A_\alpha({\bf p}) =  p_\alpha B \label{DC11b}
\end{equation}
	which turns into the usual Einstein relation in the limit $m\ll M$
\begin{equation}
	M T A_\alpha({\bf p}) =  p_\alpha B. \label{Eins_rel}
\end{equation}

	Straightforward differentiation of $W(\vec{p},\vec{q})$ in Eq.~(\ref{w_b}) with respect to
	the variable $\vec{p}\!\cdot\!\vec{q}$ which is assumed independent
	in the case of the equilibrium Maxwellian
	distribution leads to equality (\ref{DC11b}) and
	we arrive at the ordinary Fokker-Planck equation in
	velocity space with the constant diffusion coefficient $D= B /M^2$ and
	the constant friction coefficient $\beta = mB/\mu T=D(m+M)/T$ which satisfy the
	Einstein relation in the limit $m\ll M$.

	For systems close to equilibrium the PT-function is calculated and
	the appropriate Fokker-Planck equation is discussed in detail in review \cite{Hoare1971}.
   	In case of hard-sphere interactions and the equilibrium Maxwellian distribution $f_b$
	equation~(\ref{w_b}) is reduced to the equation which is consistent with the result
	given in papers \cite{Hoare1971, Nielsen1964}.
	However, even for quasi-equilibrium regimes when long tails of the PTF are
	absent the consideration in \cite{Hoare1971} is restricted to hard sphere interactions.
	Non-equilibrium forms of the PTF and changes in the structure of the Fokker-Planck
	equation for such systems are considered in \cite{Trigger 2009, Trigger 2010}.
	In the following section this consideration is extended. It brings up the importance of
	of the additional term in the expansion of the PTF with respect to the variable $p^2$
	 (see Eq.~(\ref{DC7b})). The chapter also gives correct coefficients in the generalised
	 Fokker-Planck	 equation.

\section{Generalized Fokker-Planck equation}
	In some non-equilibrium (stationary or non-stationary) systems PTF can have a long
	tail as a function of $q$. For such systems the ordinary Fokker-Planck equation~(\ref{DC3b})
	is not valid as the kinetic coefficients diverge at large values of $q$.

	Now let us refer to the expansion of the PTF given by equation~(\ref{DC7b})
	making no assumptions as for the dependence of the differential cross-section
	on the relative velocity $u$.
	It results in an additional (cp. \cite{Trigger 2009, Trigger 2010}) term proportional
	to the first derivative of the PTF with respect to $p^2$.
	Substituting this expansion into Eq.~(\ref{DC2b}) and making use of the Fourier
	transformation we arrive at the following equation
\begin{equation}
	\begin{split}
		&\frac{d \hat f(\vec{s})}{dt} =\int W(q)\left[f(\vec{p}+\vec{q})
		-f(\vec{p})\right]\frac{e^{i\vec{p}\cdot\vec{s}}}{(2\pi)^3}d\vec{p}d\vec{q}\\	
		&+\int\tilde W'(q)\left[(\vec{q}\!\cdot\!(\vec{p}+\vec{q}))f(\vec{p}+\vec{q})\right.\\
		&\left.-(\vec{q}\!\cdot\!\vec{p})f(\vec{p})\right]\frac{e^{i\vec{p}\cdot\vec{s}}}{(2\pi)^3}d\vec{p}d\vec{q}\\
		&+\int\tilde W''(q)\left[(\vec{q}\!\cdot\!(\vec{p}+\vec{q}))^2f(\vec{p}+\vec{q})\right.\\
		&\left.-(\vec{q}\!\cdot\!\vec{p})^2f(\vec{p})\right]\frac{e^{i\vec{p}\cdot\vec{s}}}{(2\pi)^3}d\vec{p}d\vec{q}\\	
		&+\int\hat W'(q)\left[(\vec{p}+\vec{q})^2f(\vec{p}+\vec{q})\right.\\
		&\left.-p^2f(\vec{p})\right]\frac{e^{i\vec{p}\cdot\vec{s}}}{(2\pi)^3}d\vec{p}d\vec{q},	
	\end{split}
\end{equation}
	where $\hat f(\vec{s})$ is the Fourier image of the function $f(\vec{p})$. We omitted
	the arguments $t$ and $\vec{q}$ of the distribution function to make the notation brief. Introducing the
	coefficients

\begin{equation}
	\begin{split}
		&A(\vec{s}) = \int \left[e^{-i{\bf(q\cdot s)}}-1\right]W(q)  d{\bf q},\\
		&B_{\alpha}(\vec{s})= -\frac{i}{s^2}\int (\vec{q}\!\cdot\!\vec{s})
			\left[e^{-i{\bf(q\cdot s)}}-1\right]\tilde W'(q)d{\bf q}\\
		&C_{\alpha\beta} (\vec{s})= -\frac{1}{2}\int  q_\alpha q_\beta
		 \left[e^{-i{\bf(q\cdot s})}-1\right]\tilde W''(q)d{\bf q},\\
		&E(\vec{s})= -\int \left[e^{-i{\bf(q\cdot s)}}-1\right]\hat W'(q)d\vec{q}
	\end{split}
\end{equation}
	we come to the Fokker-Planck-like equation
\begin{equation}
	\begin{split}
	&\frac{d\hat f({\bf s})}{dt} = A(\vec{s})\hat f ({\bf s})+ B_\alpha(\vec{s})
	\frac{\partial \hat f ({\bf s})}{\partial s_\alpha}\\
	&+C_{\alpha\beta}(\vec{s})\frac{\partial^2\hat f ({\bf s})}{\partial s_\alpha\partial s_\beta}
	+E(\vec{s}) \Delta \hat f(\vec{s}).
	\end{split}
			\label{DC16b}
\end{equation}
	After a number of calculations it can be shown that
\begin{equation}
	A(\vec{s})=A(s) = 4\pi \int_0^\infty \!\!\!q^2 \left[\frac{\sin (q s)}{qs}-1\right]W(q) dq,
			\label{DC17b}
\end{equation}

\begin{equation}
	\begin{split}
	&B_{\alpha}(\vec{s}) = s_{\alpha}B(s),\\
	&B(s)=\frac{4\pi}{s^2} \int_0^\infty \!\!\!q^2\left[\cos(q s)-\frac{\sin(q s)}{q s}\right]\tilde W'(q)dq,
	\end{split}
			\label{DC19b}
\end{equation}

\begin{equation}
	C_{\alpha\beta} (\vec{s})= s_{\alpha}s_{\beta}C_1(s)+s^2\delta_{\alpha\beta}C_2(s),
	\label{DC20b1}
\end{equation}
\begin{equation}
\begin{split}
	C_1(s) =\frac{12\pi}{s^2}\int_0^\infty q^4
		\left[\frac{\sin(qs)}{3qs}+\frac{\cos(qs)}{q^2s^2}\right.\\
		\left.-\frac{\sin(q s)}{q^3 s^3}\right]
		\tilde W''(q)dq,
		\end{split}
		\label{DC20b2}
\end{equation}
\begin{equation}
	\begin{split}
	C_2(s) =\frac{4\pi}{s^2}\int_0^{\infty}q^4\left[-\frac{1}{3}-\frac{\cos(qs)}{q^2s^2}\right.\\
		\left.+\frac{\sin(qs)}{q^3s^3}\right]
		\tilde W''(q)dq,
	\end{split}
				\label{DC20b3}
\end{equation}	

\begin{equation}
	E(\vec{s})=s^2C_3(s),
\end{equation}
\begin{equation}
	C_3(s) =  \frac{4\pi}{s^2} \int_0^\infty \!\!\!q^2 \left[1-\frac{\sin (q s)}{qs}\right]\hat W'(q) dq.
\end{equation}

	For an isotropic function $f({\bf s})\equiv f(s)$ the Fokker-Planck-like equation~(\ref{DC16b}) is 	reduced to equation
\begin{equation}
	\begin{split}
	&\frac{d\hat f(s)}{dt} = A (s)\hat f(s)\\
	&+ \left[B(s)+2C_2(s)+2C_3(s)\right]s\frac{\partial\hat f(s)}{\partial s}\\
	&+\left[C_1(s)+C_2(s)+C_3(s)\right] \frac{\partial^2 \hat f(s)}{\partial s^2}.
	\end{split}
			\label{DC22b}
\end{equation}
	This equation is a generalisation of the usual Fokker-Planck equation which implies certain
	smallness of the transferred momentum $q$. The advantage of our equation is that it is valid
	for an arbitrary value of $q$. This virtue lets us go far beyond the scope of the phenomena
	described by the ordinary Fokker-Planck equation.
	
	However, it is instructive to match Eq.~(\ref{DC22b}) and the usual Fokker-Planck equation.
	The latter tackles the problem of diffusion of heavy particle in a gas of light particles.
	Basically, Eq.~(\ref{DC22b}) should be reduced to the Fokker-Planck equation in the
	limit $m/M\rightarrow 0$ and for small values of $q$.
	First of all, let us note that in the case when the PTF and the functions $W(q)$, $\tilde W'(q)$,
	$\hat W'(q)$ and $\tilde W''(q)$ strongly decrease for large values of $q$ the
	exponents in the integrals in the functions $A(s)$, $B(s)$,
	$C_1(s)$, $C_2(s)$ and $C_3(s)$ can be expanded into a series as follows
\begin{equation}
	\begin{split}
&	A(s)\simeq - \frac{s^2}{6}\int q^2 W(q)d\vec{q},\\
&	B(s)\simeq - \frac{1}{3} \int  q^2 \tilde W'(q)d\vec{q},\\
&	C_1(s)\simeq -\frac{1}{15} \int q^4 \tilde W''(q)d\vec{q}, \\
&	C_2(s)\simeq -\frac{1}{30} \int q^4 \tilde W''(q)d\vec{q},\\
&	C_3(s)\simeq  \frac{1}{6}\int q^2 \hat W'(q)d\vec{q}.
	\end{split}
			\label{DC23b}
\end{equation}
	Then we can set the coefficients $C_1$, $C_2$ and $C_3$ equal to zero as
	they are of next order of smallness compared to the coefficients $A$ and $B$
	with respect to the small parameter $m/M$. The fact can be deduced from
	the expression for the PTF~(\ref{DC13z}).
	And the simplified kinetic equation in velocity space based on the
	PTF (which is non-equilibrium in the general case) yields

\begin{equation}
	\frac{df(s,t)}{dt} = A_0 s^2 f(s)+ Bs\PDEV{f(s)}{s},
				\label{DC25b}
\end{equation}
	where $A_0= -1/6 \int q^2 W(q)d\vec{q}$.
	
	The stationary solution of this equation is given by
\begin{equation}
	f(s) ={\rm Const}\cdot \exp\left[-\frac{A_0 s^2}{2 B}\right].
\end{equation}	
	The respective normalized stationary momentum distribution reads
\begin{equation}
	f(p) = \frac{N B^{3/2}}{(2 \pi A_0)^{3/2}}\exp\left[-\frac{B p^2}{2 A_0}\right].
				\label{DC28b}
\end{equation}
	
\section{Special cases of anomalous diffusion in velocity space}

	Let us calculate the kinetic coefficients
	in the generalized Fokker-Planck equation
	for the special cases of anomalous
	diffusion. In our model diffusing heavy particles with mass $M$ interact with
	light particles of the surrounding medium with mass $m\ll M$
	and with power-type velocity distribution
\begin{equation}
	f_b(\vec{u}+\vec{v})=\frac{n_b}{u_0^{3-2\gamma}({\vec{u}}+\vec{v})^{2\gamma}},
	  \label{power_type_dist}
\end{equation}
	where $u_0$ means the characteristic velocity,
	$n_b$ is the constant which has the dimension of the density in the coordinate space
	but since the function $f_b$ can not be normalized
	we can not think of $n_b$ as of a density.
	Impossibility to normalize the distribution function $f_b$
	is explained with the fact that the normalizing integral diverges
	either at small or at large values of the argument of $f_b$.
	In reality normalization can always be fulfilled. At smal values of the argument
    	 the power-law~(\ref{power_type_dist}) of the distribution function is violated.
	 But it is highly unlikely to obtain an analytical expression for $W ({\bf q, p})$ for realistic
    	distribution functions.
	
\subsection{Hard-sphere interactions}

	Interactions in systems of hard-sphere particles of radius $a$ are
	described by the differential cross-section $d\sigma/do=a^2/4$.
	Substitution of the specified distribution function
	and cross-section into equation~(\ref{w_b}) gives
\begin{equation}
	W ({\bf q, p})= \frac{n_b a^2}{4\mu^2 q u_0^{3-2\gamma}}
	\int \frac{\delta\!\left(u_\rmz+\frac{q}{2\mu}\right)}{(\vec{u}+\vec{v})^{2\gamma}}
	d^3\vec{u}.
		\label{DC30b}
\end{equation}

	For $\gamma>1$ the integral~(\ref{DC30b}) converges and the result
	of integration reads
\begin{equation}
	W(\vec{p},\vec{q}) =
	\frac{\pi n_b a^2(\xi-1)^{2-2\gamma}}
	{(2\mu)^{4-2\gamma}(\gamma-1)q^{2\gamma-1}u_0^{3-2\gamma}},
		\label{w_b_hard}
\end{equation}
	where $\xi = 2\mu(\vec{q}\cdot\vec{v})/q^2$. Expanding equation~(\ref{w_b_hard})
	into a series according to~(\ref{DC7b}) we leave only the first three terms
\begin{equation}
	W(q) = \frac{\pi n_b a^2}{(2\mu)^{4-2\gamma}(\gamma-1)q^{2\gamma-1}u_0^{3-2\gamma}}
	     = \frac{R_0}{q^{2\gamma-1}},
\end{equation}
\begin{equation}
\begin{split}
	&\tilde{W}'(q) = \frac{2\pi n_b a^2}{M(2\mu)^{3-2\gamma}q^{2\gamma+1}u_0^{3-2\gamma}}
	     = \frac{R_1}{q^{2\gamma+1}}, \\
	&R_1 = \frac{2\mu(2\gamma-2)}{M}R_0,
	  \end{split}
\end{equation}
\begin{equation}
\begin{split}
	&\tilde{W}''(q) = \frac{2\pi n_b a^2(2\gamma-1)}
	{M^2(2\mu)^{2-2\gamma}q^{2\gamma+3}u_0^{3-2\gamma}}
	     = \frac{R_2}{q^{2\gamma+3}}, \\
	&R_2 =  \frac{(2\mu)^2(2\gamma-2)(2\gamma-1)}{M^2}R_0,
\end{split}
\end{equation}
\begin{equation}
	\hat W'(q) = 0.
\end{equation}

	Now let us substitute $W(q)$ into equation~(\ref{DC17b}) which introduces the coefficient $A(s)$
	and discuss the convergence criteria.
\begin{equation}
\begin{split}
	A(s) = 4\pi \int_0^\infty  q^2 \left[\frac{\sin\,(qs)}{qs}-1\right]W(q)dq\\
	     = 4\pi R_0 \int_0^\infty\frac{dq}{q^{2\gamma-3}}\left[\frac{\sin\,(q s)}{q s}-1\right]
			\label{DC33b}
\end{split}
\end{equation}

	The convergence of the integral in equation~(\ref{DC33b}) is guaranteed when the inequality $2<\gamma<3$
	takes place. Whereas $\gamma<3$ gives the
	convergence for small values of $q$ ($q\rightarrow0$) the convergence for $q\rightarrow\infty$
	is provided at $\gamma>2$.

	The values of the parameter $\gamma$ at which the coefficient $B(s)$ remains finite can be found
	directly from the definition of $B(s)$ (Eq.~(\ref{DC19b})).
	It is clear to see that condition for convergence
	of the integral in equation for the coefficient $B(s)$
	for small values of $q$ is $\gamma<2$ and for
	large values of $q$ is $\gamma>1/2$.
	Finally, finiteness of the coefficients $C_1(s)$ and $C_2(s)$ follows from the inequality
	$\gamma<2$ for small values of $q$. For large values of $q$ the coefficient $C_1$
	is finite at $\gamma>1$ and the coefficient $C_2$ is finite at $\gamma>0$ (Eq.~(\ref{DC20b1}-(\ref{DC20b3})).

	To sum up, in the case under consideration, the convergence of the coefficients $A$, $B$, and $C_{1,2}$ for
	large values of $q$ is defined only by the convergence of the coefficient $A$, which means $\gamma>2$.
	For small values of $q$ it is sufficient to guarantee
	convergence for the coefficient $B(s)$ with $\gamma<2$.
	In other words, for purely power-type behavior of the function $f_b(\xi)$
	the simultaneous convergence of the coefficients $A$, $B$ and $C_{1,2}$ does not exist.
	However, in real physical models, the convergence of the
	coefficients at $q\rightarrow 0$ can be achieved, for example,
	due to the non-power-type behavior of the PTF $W$ for small values of $q$
	(compare with examples of anomalous diffusion in coordinate space \cite{Trigger2005}).
	Therefore, in the system of hard-sphere heavy particles undergoing anomalous diffusion
	in the medium of light particles with power-type distribution
	at large values of the variable $q$ the kinetic coefficients exist at
	$\gamma>2$.
	This result approves of the qualitative conclusion in paper
	\cite{Trigger 2009} but the numbers do not coincide as we started from another more explicit
	expression for the PTF (\ref{DC13z}).

\subsection{Coulomb collisions}
	By analogy, we can consider a system of particles characterized by Coulomb interactions.
	Formally, the Coulomb interactions are described by a differential cross-section
\begin{equation}
	\frac{d\sigma_{\rm Coul}}{do}  = \left(\frac{Ze^2}{2\mu u^2}\right)^2\frac{1}{\sin^4\!\chi/2}
		= \frac{4Z^2e^4\mu^2}{q^4},
			\label{Coul_sigma}
\end{equation}
	where $Z$ is the charge number and the scattering angle
	$\chi = \arccos(1-q^2/2\mu u^2)$ (see Eq.~(\ref{w_b})).
	From the mathematical point of view analysis of the kinetic coefficients in the
	case of Coulomb collisions is no more difficult than the case of hard-sphere collisions
	as the Coulomb cross-section is also independent on the relative velocity $u$.
	The only difference is the power in the dependence of the PTF on the transfer momentum $q$
\begin{equation}
	W_{\rm Coul} ({\bf q, p})= \frac{4Z^2e^4n_b}{q^5 u_0^{3-2\gamma}}
	\int \frac{\delta\!\left(u_\rmz+\frac{q}{2\mu}\right)}{(\vec{u}+\vec{v})^{2\gamma}}
	d^3\vec{u}.
		\label{W_coul}
\end{equation}
Performing integration we arrive at the equation
\begin{equation}
	W_{\rm Coul}(\vec{p},\vec{q}) = \frac{4\pi Z^2e^4n_b(\xi-1)^{2-2\gamma} }
	{(2\mu)^{2-2\gamma}(\gamma-1)q^{2\gamma+3}u_0^{3-2\gamma}},
		\label{w_b_coul}
\end{equation}
Expanding the PTF into a series we obtain the functions $W_{\rm Coul}(q)$, $\tilde{W}'_{\rm Coul}(q)$ and
$\tilde{W}''_{\rm Coul}(q)$
\begin{equation}
\begin{split}
	W_{\rm Coul}(q) = \frac{4\pi Z^2e^4 n_bu_0^{2\gamma-3} }
	{(2\mu)^{2-2\gamma}(\gamma-1)q^{2\gamma+3}}
	     = \frac{K_0}{q^{2\gamma+3}},
\end{split}
\end{equation}
\begin{equation}
\begin{split}
	&\tilde{W}_{\rm Coul}'(q) = \frac{8\pi Z^2 e^4 n_bu_0^{2\gamma-3} }
	{M(2\mu)^{1-2\gamma}q^{2\gamma+5}}
	     = \frac{K_1}{q^{2\gamma+5}}, \\
	&K_1 = \frac{2\mu(2\gamma-2)}{M}K_0,
\end{split}
\end{equation}

\begin{equation}
\begin{split}
	&\tilde{W}_{\rm Coul}''(q) = \frac{8\pi Z^2e^4 n_b(2\gamma-1)u_0^{2\gamma-3} }
	{M^2(2\mu)^{-2\gamma}q^{2\gamma+7}}
	     = \frac{K_2}{q^{2\gamma+7}}, \\
	&K_2 =  \frac{(2\mu)^2(2\gamma-2)(2\gamma-1)}{M^2}K_0,
\end{split}
\end{equation}
\begin{equation}
	\hat W_{\rm Coul}'(q) = 0.
\end{equation}

	Reasoning by analogy with the previous section concerning hard-sphere interactions
	we can draw a conclusion
	that the coefficients $A(s)$, $B(s)$ and $C_{1,2}(s)$ are finite at large values of $q$
	provided the inequality
	$\gamma>0$ is fulfilled. We assume that the convergence of the integrals in these
	 coefficients at small
	values of $q$ is due to various neglected in our model factors, such as screening
	effect, etc.
    	The asymptotic behavior of the PTF and the respective derivatives of the PTF for
	 large $q$ is similar to one obtained in \cite{TrEbHeScSok 2010}.
	 However, the coefficients are calculated on the basis of the explicit representation
	  for the PTF (\ref{DC13z}).

\subsection{Interactions typical for dusty plasmas}

	A more complicated and interesting case is interaction of dusty particles
	with electrons and ions.
	The differential cross-section for this type of interaction depends on the relative velocity
	of the particles unlike the cases of Coulomb and hard-sphere models of collisions.
	This fact leads to the dependence of the PTF on $p^2$. To demonstrate this let us refer to the
	the differential cross-section for grains-electrons interactions \cite{Fortov2004}.
	The respective differential cross-section yields
\begin{equation}
	\sigma_{\rm e}(u) = \left\{
	\begin{split}
&		\pi \rho^2\left(1-\frac{2e^2}{\rho m_{\rm e} u^2}\right),
		\quad\frac{2e^2}{\rho m_{\rm e} u^2}<1,\\
&		0,\quad \frac{2e^2}{\rho m_{\rm e} u^2}>1,
	\end{split}
	\right.
		\label{dust_cross_section}
\end{equation}

\begin{equation}
	\sigma_{\rm i}(u)=\pi \rho^2\left(1+\frac{2e^2}{\rho m_{\rm i} u^2}\right),
\end{equation}
where $m_{\rme(\rmi)}$ is electron (ion) mass, $\rho$ is radius of a grain,
$e$ is the elementary charge. The ions are assumed singly charged.

	Let us consider electron-grain collisions and rewrite Eq.~(\ref{dust_cross_section}) in a more
	convenient way
\begin{equation}
	\sigma_{\rm e}(u) = \left\{
	\begin{split}
&		\frac{\rho^2}{4}\left(1-\frac{\Delta^2}{u^2}\right),
		\quad u>\Delta,\\
&		0,\quad u<\Delta,
	\end{split}
	\right.	\label{Dust_cross-section1}
\end{equation}
	where $\Delta^2 = 2e^2/\rho m_\rme$. Along with the previous special cases we assumed	
	scattering particles (electrons) have a power-type velocity distribution~(\ref{power_type_dist}).
	Substituting the announced cross-section and the distribution function into the PTF~(\ref{w_b})
	we arrive at
\begin{equation}
\begin{split}
	&W(\vec{q},\vec{v}) = \frac{\rho^2n_b}{4\mu^2u_0^{3-2\gamma}q}
			    \int_{u>\Delta}\left(1-\frac{\Delta^2}{u^2}\right)\\
	&	\times	    \delta\left(u_\rmz+\frac{q}{2\mu}\right)
			    \frac{d\vec{u}}{(\vec{u}+\vec{v})^{2\gamma}}.
\end{split}
\end{equation}
	
	Analyzing the integral we can conclude that for large values of $q/2\mu>\Delta$
	the inequality $u>\Delta$ is guaranteed. In our paper we restrict our consideration only
	with large values of $q$. The point is that the power-type distribution function
	is realistic only for large values of $q$. For small values it diverges and
	the processes which suppress its growth should be taken into account.
	
	But even for $q/2\mu>\Delta$ when the integration should be performed
	over the whole velocity space it is not possible to obtain an analytical expression
	for the PTF in terms of elementary functions.
	However, we can calculate the coefficients $W(q)$, $\tilde W'(q)$, $\tilde W''(q)$
	and $\hat W'(q)$ expanding the integral into a series in the vicinity of the point
	$\SCAL{q}{p}=0$ a $p^2 = 0$ when it is necessary.
\begin{equation}
\begin{split}
&	W(q) = \frac{\pi\rho^2}{4\mu^2q} \frac{n_bu_0^{2\gamma-3}}{\bar q^{2(\gamma+1)}}\\		&		\times\left[\frac{1}{\gamma+1}-\frac{\Delta^2}{\bar q^2}\frac{1}{\gamma+2}\right]\!,
\end{split}
\end{equation}
	where $\bar q = q/2\mu$.
\begin{equation}
\begin{split}
&	\hat W'(q) = \frac{\pi\rho^2}{4M^2\mu^2q}\frac{\gamma}{\gamma+3}
	\frac{n_bu_0^{2\gamma-3}}{\bar q^{2(\gamma+2)}}\\
&	\times \left[\frac{\Delta^2}{\bar q^2}\frac{3}{\gamma+4}-\frac{2}{\gamma+2}\right].
\end{split}
\end{equation}
	\begin{equation}
\begin{split}
&	\tilde W'(q) = \frac{\pi\rho^2}{M\mu q}\frac{n_bu_0^{2\gamma-3}}{\bar q^{2(\gamma+2)}}\\
&		\times\left[\frac{1}{\gamma+2}-\frac{\Delta^2}{\bar q^2}\frac{1}{\gamma+3}\right].
\end{split}
\end{equation}
\begin{equation}
\begin{split}
&	\tilde W''(q) = \frac{2\pi\rho^2}{M^2\mu^2q}\frac{\gamma(\gamma+1)}{\gamma+3}
	\frac{n_bu_0^{2\gamma-3}}{\bar q^{2(\gamma+3)}}\\
&	\times\left[\frac{\Delta^2}{\bar q^2}\frac{1}{\gamma+4}-\frac{1}{\gamma+2}\right].
\end{split}
\end{equation}
	The convergences of all coefficients above at large values of $q$ fulfills when $\gamma>0$.
	Remarkably for the dusty particles interacting with electrons the coefficient $\hat W'(q)$
	is not equal to zero although it is small due to the mass ratio of electrons and ions.

	Similar analysis can be done for the dusty particles interacting with ions.

\section{Conclusion}
A more general Fokker-Planck-like equation is derived on the basis of the master equation. The equation is applicable for any values of the transferred momentum in a collision act unlike the usual Fokker-Planck equation which is valid for relatively small values of the transferred momentum. The mass ratio of the interacting particles is not important for the derived Fokker-Planck-like equation. It is based on the mere assumption that there are two species of the colliding
particles, namely, with large and small velocity values.

The coefficients for the general Fokker-Planck-like equation are calculated using the general expression for the probability transition function describing elastic collisions. The scattering particles were assumed to have a velocity distribution function with power-type tail. It leads to the power-type
dependence of the PTF on the transferred momentum and the corresponding process of diffusion  in velocity space has anomalous character. Three different examples of the differential cross-section have been considered. The hard-sphere model of collisions as well Coulomb model
demonstrate similar behavior of the PTF. However, for Coulomb interactions PTF has a more strongly decreasing tail. The case of dusty particles scattering on plasma particles (electrons or ions) is more complicated leads to the less trivial dependence of the PTF on the transferred momentum.

\section{Acknowledgements}
The work is supported by the Project
RFBR 10-02-90418-${\rm Ukr}\_a$ and by Non-Profit Foundation Dynasty.

\end{document}